\documentclass[a4paper,preprint,aps,prb,showpacs,showkeys,superscriptaddress]{revtex4-1}

\setcitestyle{super}
\usepackage[utf8]{inputenc}
\usepackage{amsmath,amsfonts,amssymb}
\usepackage{graphicx}
\usepackage{dcolumn,subfigure}
\usepackage{textcomp,array,url,color} 
\usepackage{multirow,booktabs,bm}
\usepackage[squaren,Gray,cdot]{SIunits}
\usepackage{xspace,rotating,tabularx}

\newcolumntype{Y}{>{\centering\arraybackslash}X}
\newcommand{\ie}{\textit{i.e.}\@\xspace}

\newcommand{\etal}{\emph{et al.~}}

\usepackage[pdfstartview=FitV,bookmarks=true,
            linktocpage=true,colorlinks=true,
            urlcolor=blue,linkcolor=red,
            citecolor=blue]{hyperref}

\begin{document}

%\preprint{Draft ZGV --- TI cylinder --- JASA}

\title{Laser induced Zero-Group Velocity resonances in Transversely Isotropic cylinder}

\author{J\'{e}r\^{o}me Laurent}%\email{jerome.laurent@espci.fr}
\affiliation{Institut Langevin, ESPCI ParisTech, PSL Research University,\\ CNRS UMR 7587, 1 rue Jussieu, 75005 Paris, France}
\author{Daniel Royer}%\email{daniel.royer@espci.fr}
\affiliation{Institut Langevin, ESPCI ParisTech, PSL Research University,\\ CNRS UMR 7587, 1 rue Jussieu, 75005 Paris, France}
\author{Takasar Hussain}%\email{htakasarnust@gmail.com}
\affiliation{School of Natural Sciences, National University of Sciences and Technology,\\ Sector H-12, Islamabad, Pakistan}
\author{Faiz Ahmad}%\email{faizmath102@gmail.com}
\affiliation{School of Natural Sciences, National University of Sciences and Technology,\\ Sector H-12, Islamabad, Pakistan}
\author{Claire Prada}%\email{claire.prada@espci.fr}
\affiliation{Institut Langevin, ESPCI ParisTech, PSL Research University,\\ CNRS UMR 7587, 1 rue Jussieu, 75005 Paris, France}

\pacs{43.20.Bi, 43.35.Cg, 43.20.Ks, 43.35.Zc}
% 43.20.Bi  -->  Mathematical theory of wave propagation
% 43.20.Mv  -->  Waveguides, wave propagation in tubes and ducts
% 43.35.Cg  -->  Ultrasonic velocity, dispersion, scattering, diffraction, and attenuation in solids; elastic constants
% 43.20.Ks  -->  Standing waves, resonance, normal modes
% 43.25.Gf  -->  Standing waves; resonance
% 43.35.Yb  -->  Ultrasonic instrumentation and measurement techniques
% 43.35.Zc  -->  Use of ultrasonics in nondestructive testing, industrial processes, and industrial products
% 43.40.Cw  -->  Vibrations of strings, rods, and beams
% 43.40.Fz  -->  Acoustic scattering by elastic structures
\keywords{Laser-Ultrasonic technique, Zero-Group Velocity modes, TI cylinder}
\date{\today}

\begin{abstract}
The transient response of an elastic cylinder to a laser impact is studied. When the laser source is a line perpendicular to the cylinder axis, modes guided along the cylinder are generated. For a millimetric steel cylinder up to ten narrow resonances can be locally detected by laser interferometry below 8 MHz. Most of these resonances correspond to Zero-Group Velocity guided modes while a few others can be ascribed to thickness modes. We observe that the theory describing the propagation of elastic waves in an isotropic cylinder is not sufficient to precisely predict the resonance spectrum. In fact, the texture of such elongated structure manifest as elastic anisotropy. Thus, a transverse isotropic (TI) model is used to calculate the dispersion curves and compare them with the measured one, obtained by moving the source along the cylinder. The five elastic constants of a TI cylinder are adjusted leading to a good agreement between measured and theoretical dispersion curves. Then, all the resonance frequencies are satisfactorily identified. 
\end{abstract}

\maketitle

%           Introduction
% ------------------------------------
\section{Introduction}

Elongated cylindrical structures like rods, cable strands or fibers are widely used in the industry or in civil engineering. Characterization of mechanical properties of constitutive materials is important for testing their structural integrity. Non Destructive Evaluation (NDE) of these properties is usually carried out with elastic waves. Various methods like Resonant Ultrasonic Spectroscopy (RUS)/pulse-echo contact-method, or pulsed/continuous laser-ultrasound (LU) contactless-method are used. Cylindrical structures support the propagation along their axes of elastic waves of different types: longitudinal (L), flexural (F) or torsional (T). Many theoretical and numerical studies and few experimental results were published this past fifty years on the propagation of time harmonic guided modes in solid or hollow cylinders.\citep{Gazis59,Mindlin60,Meeker64} In the nineteenth century, Pochhammer and Chree first established the equation of longitudinal waves in free isotropic cylindrical structures. In 1943, Hudson studied the fundamental flexural mode in a solid cylinder~\cite{Hudson43} and longitudinal modes of a bar were examined by Davies~\cite{Davies48} in 1948. Gazis reported the first exact solutions of the frequency equation,\cite{Gazis59} as well as a complete description of propagative modes, displacement and stress distributions for an isotropic elastic hollow cylinder in vacuum. Pao and Mindlin~\citep{Pao60,Pao62} as well as Onoe, McNiven and Mindlin~\cite{Onoe62} studied all the branches of the complete three dimensional problem of a free solid cylinder. A thorough review of elastic wave propagation in isotropic elastic cylinders and plates was given by Meeker and Meitzler.\cite{Meeker64} Later, Zemanek investigated elastic wave propagation in a cylinder, both experimentally and theoretically.\cite{Zemanek72}

All these works deal with isotropic material, however transverse anisotropy is exhibited by elongated cylindrical structures due to their manufacturing processes (poly-crystalline metals) \cite{Mason99} or their texture (carbon fibers used in reinforced polymers). Then, non-destructive measurement of elastic constants of transversely isotropic (TI) materials is of great interest,\cite{Payton83} especially in aeronautic and aerospace industries. Morse first established the frequency equation for longitudinal waves propagating in TI cylinders.\cite{Morse54} The extension of Gazis formulation to orthotropic and TI waveguides was initiated by Mirsky in 1964 for infinite and finite cylinders.\citep{Mirsky64,Mirsky65a} Then, several researchers investigated the propagation or the scattering of elastic waves in free or fluid-loaded TI cylinders.\citep{Eliot68,Mcniven71,Nagy95,Berliner96a,Niklasson98} Ahmad and Rahman~\cite{Ahmad00} also studied the scattering of acoustic wave incident on a TI cylinder and showed that Buchwald's representation~\cite{Buchwald61} yields much simpler equations.\cite{Ahmad01} This representation is beneficial to simplify the description of the potential functions and economizes laborious calculations. This model provides perfectly similar results with Honarvar and Sinclair model~\cite{Honarvar96,Honarvar07} and can be applied to study both isotropic and TI cylinders. Honarvar \etal also obtained the frequency equations of axisymmetric and asymmetric free vibrations of finite TI cylinders.\cite{Honarvar09} Several other researchers proposed to use the impedance matrix theory~\cite{Norris10,Norris13} or a SAFE method to study transient thermoelastic waves in isotropic/anisotropic cylinders.\cite{Chitikireddy11}\\

Frequency equations determining the angular frequency $\omega(=2\pi f)$ versus the axial wave number $k(=2\pi/\lambda)$ have to be solved numerically. For a given circumferential order $n$, the various solutions (integer $m$) can be grouped into different families: longitudinal L$(0,m)$ and torsional T$(0,m)$ modes with displacements independent of the azimuthal angle $\theta$, or flexural F$(n,m)$ modes with displacements varying as $\sin n\theta$ or $\cos n\theta$. Each mode is represented by a dispersion curve $\omega(k)$. Many similarities exist between the families of modes found in plates and cylinders. Indeed, torsional modes T$(0,m)$ are similar to shear modes SH in plates. Longitudinal L$(0,m)$ and first flexural F$(1,m)$ modes in cylinders are analogous to symmetrical $\text{S}_m$ and anti-symmetrical $\text{A}_m$ modes in plates. Major differences appear in the frequency spectrum for flexural modes F$(n,m)$ with family number $n\geqslant2$.\\

Laser-based ultrasonic technique is a convenient tool for the generation and detection of guided elastic waves in plates \cite{Hutchins89} or cylinders.\citep{Gsell04} It has been shown that this non-contact technique is very efficient to observe Zero Group Velocity (ZGV) Lamb modes corresponding to a frequency minimum of the dispersion curves $\omega(k)$.\citep{Prada05a,Prada08b} For these specific points, the energy deposited by the laser pulse remains trapped under the source. The resulting local and narrow resonances can be detected at the epicenter with an optical interferometer. This non-contact method allowed us to perform accurate measurements of elastic properties of isotropic or transverse isotropic materials.\citep{Clorennec07,Ces12} For cylinders, a line laser source can be used to control the propagation direction. For a line source parallel to the cylinder axis, circumferential Rayleigh and whispering gallery waves are excited giving rise to resonances at all $n$ integer values of the normalized circumferential wavenumber~\cite{Clorennec02} $ka$. This configuration was also used by Mounier \etal to study resonances of a micrometric fiber at sub-gigahertz frequencies.\citep{Mounier14a,Mounier14b}\\
 
In this paper, we investigate the mechanical response of a transversely isotropic cylinder to a laser line source perpendicular to the axis. In this case, axially guided elastic waves are generated and many resonances are observed. We performed measurements in a millimeter steel cylinder. Many resonances are observed in the MHz range. They can be roughly identified from the minimum frequency of some branches of the dispersion curves of longitudinal and flexural modes calculated for an isotropic cylinder. However, significant discrepancies remain between the isotropic model and experimental resonance frequencies. Dispersion curves are then calculated using a transverse isotropic model. The five independent elastic constants are adjusted such that all measured resonance frequencies can be precisely predicted. Moreover, experimental dispersion curves, measured by moving the laser source, are compared with the theoretical ones.

%      Experimental results
% ------------------------------------
\section{Experiments}\label{sec:exp}

Measurements were performed with an optical interferometer at the center of the source with the laser line perpendicular to the cylinder axis. Resonances were extracted from the spectrum of the temporal signal. Their frequencies were compared with minimum frequencies of dispersion curves calculated from an isotropic material.

\subsection{Laser-Ultrasonic setup}

\begin{figure}[!ht]
\includegraphics[width=0.55\columnwidth]{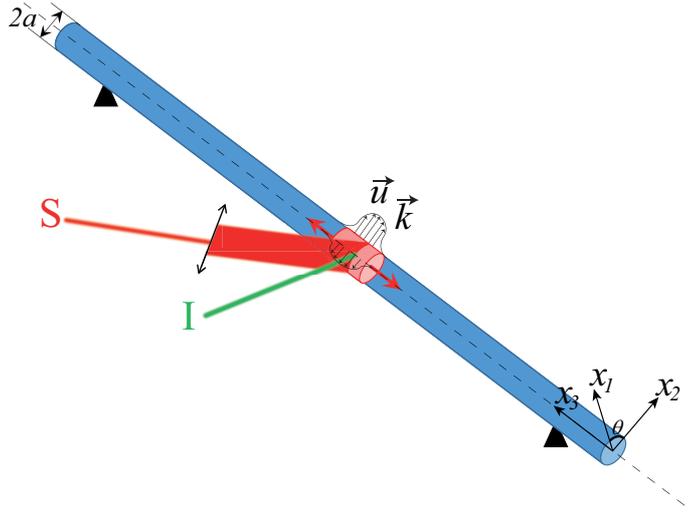} 
\caption{Source and probe geometry used to excite and to detect local resonances in a cylinder. S: laser source, I: interferometer, C: cylindrical lens. (color online)}
\label{setup}
\end{figure}

The experimental setup is shown Fig.~\ref{setup}. The sample is an austenitic stainless steel (AISI 304L) solid cylinder (length $420$ mm, diameter $2a=0.775$ mm). The longitudinal and transverse velocities are $V_L$ = 5650 m/s and $V_T$ = 3010 m/s respectively, and the measured mass density is $\rho=$ \unit{7.91\times 10^3}{\kilogram.\meter\rpcubed}. The rod is supported by two bevelled metallic pieces to reduce the mechanical contact (friction). Elastic waves were generated by a Q-switched Nd:YAG (yttrium aluminium garnet) laser providing 8-ns pulses of 15-mJ energy at a 100-Hz repetition rate (Quantel Centurion). The spot diameter of the unfocused beam is equal to 2.5 mm. A beam expander ($\times 4$) and a cylindrical lens (focal length 250 mm) were used to enlarge and focus the laser beam into a narrow line on the surface. The optical energy distribution was close to a Gaussian and the absorbed power density was below the ablation threshold. The full length of the source at $1/e$ of the maximum value was found to be 10 mm and the width was estimated to be 0.3 mm. In the thermoelastic regime, the source is equivalent to a set of force dipoles distributed on the surface perpendicularly to the line.\\

Local vibrations were measured by a heterodyne interferometer equipped with a 100-mW frequency doubled Nd:YAG laser (optical wavelength $\Lambda$= 532 nm). This interferometer is sensitive to any phase shift $\Delta\phi$ along the path of the optical probe beam, and then to the mechanical displacement $u_r$ normal to the surface. The calibration factor (85 nm/V), deduced from the phase modulation $\Delta\phi = 2\pi u_r/\Lambda$ of the reflected beam, was constant over the detection bandwidth (20 kHz - 20 MHz). Large low frequency phase-shifts due to thermal effect or first flexural F(1,1) mode are eliminated by interposing, before amplification, a high-pass filter having a cut-off frequency equal to 1.5 MHz. Measurements were conducted at room temperature ($21\pm 0.5\celsius$). Signals detected by the optical probe were fed into a digital sampling oscilloscope and transferred to a computer.

\subsection{Zero-Group Velocity resonances}

\begin{figure}[!ht]
\subfigure{\includegraphics[width=0.9\columnwidth]{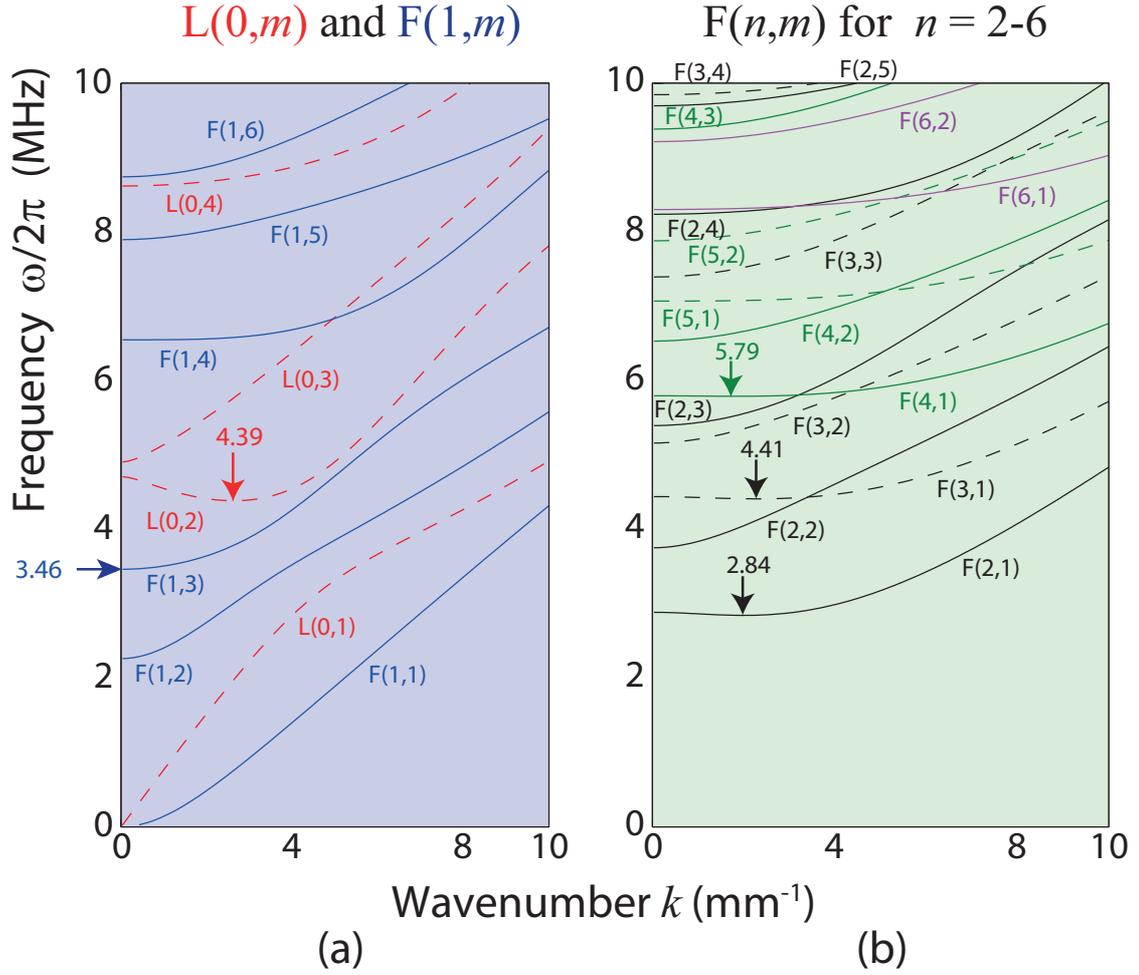}\label{disp_curve_iso1}}
\subfigure{\label{disp_curve_iso2}}
\caption{Dispersion curves for an isotropic stainless steel solid cylinder (diameter 0.775 mm). a) Longitudinal L$(0,m)$ and first flexural F$(1,m)$ modes. b) Higher order flexural modes F$(n,m)$. Minimum frequencies corresponding to ZGV modes are indicated by vertical arrows. (color online)}
\label{disp_curve_iso}
\end{figure}

The propagation of elastic waves along the cylinder axis is represented by dispersion curves calculated from Zemanek equation.\cite{Zemanek72} The numerical algorithm used to find the roots of the secular equation was proposed by Seco and Jiménez.\cite{Seco12} Firstly, the cut-off frequencies are evaluated with a bisection method, then a zero-finding algorithm is applied to determine each branch successively. The roots were obtained with an acceptable error of less than $10^{-6}$. Finally, the dispersion curves are determined for longitudinal modes and the first seven families of flexural modes in less than one minute on a personal computer, with a resolution $\Delta k=10^{-3}$ \reciprocal{\milli\meter}. As shown in Fig.~\ref{disp_curve_iso1}, dispersion curves obtained for longitudinal L$(0,m)$ and first flexural F$(1,m)$ modes are similar to those obtained for an isotropic elastic plate. Dispersion curves of higher order flexural modes F$(n,m)$ are plotted in Fig.~\ref{disp_curve_iso2}. As the interferometer is only sensitive to the normal displacement, torsional modes are not presented.\\

\begin{figure}[!ht]
\subfigure{\includegraphics[width=0.8\columnwidth]{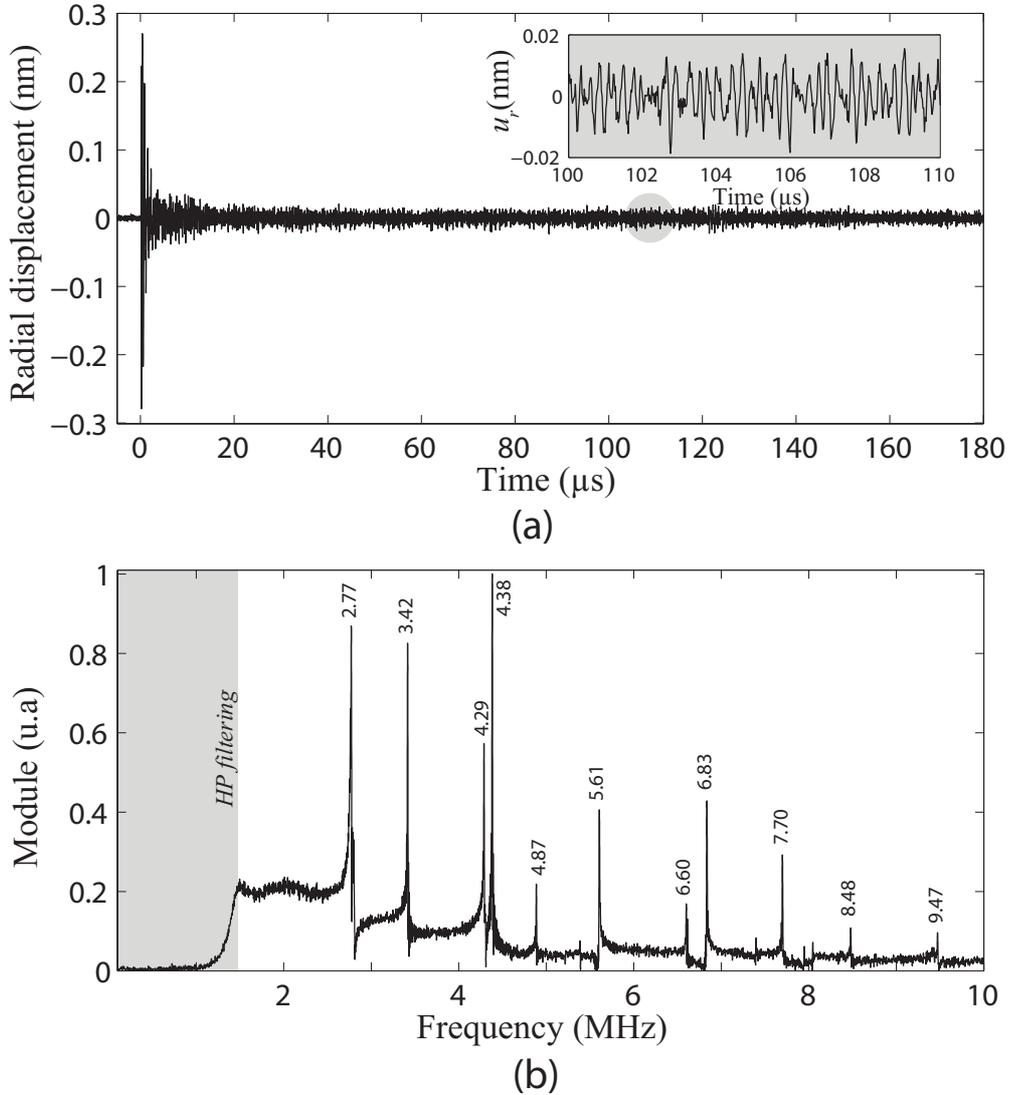}\label{Signal_Spectre1}}
\subfigure{\label{Signal_Spectre2}}
\caption{a) Radial displacement measured by the optical interferometer. Inset: zoom on a portion of the temporal displacement. b) Frequency spectrum filtered by an analog high-pass filter ($f\leqslant1.5$ MHz). (color online)}
\label{Signal_Spectre}
\end{figure}

A typical signal, corresponding to the mechanical displacement normal to the cylinder surface, is given in Fig.~\ref{Signal_Spectre1}. As previously explained, the oscillations in the first 5 µs are due to the large displacements associated to the low frequency components of the flexural mode F$(1,1)$ similar to the $\text{A}_\text{0}$ mode in plates. As shown in the insert, the low amplitude tail for $t>5$ µs is not noise but high frequency oscillations due to the ZGV resonances. The spectra of out-off plane displacement are shown in Fig.~\ref{Signal_Spectre2}. Seven resonances dominate between 0 and 8 MHz. The peak at 2.77 MHz can be ascribed to the ZGV resonance at the minimum frequency of the F$(2,1)$-mode, while peaks at 4.38 and 4.29 MHz correspond to L$(0,2)$ and F$(3,1)$-ZGV modes, respectively. The peak at 3.42 MHz can be associated with a thickness resonance [horizontal arrow in Fig.~\ref{disp_curve_iso1}] at the cut-off frequency (3.46 MHz) of the F$(1,3)$ mode. The peak at 5.61 MHz is relatively close to the minimum frequency (5.79 MHz) of the F$(4,1)$ mode. Higher frequency peaks at 6.83 and 7.70 MHz do not correspond to any minimum frequency on the dispersion curves. Except for L(0,2) which is estimated at $0.2\%$, relative errors from $2\%$ to $4\%$ are observed for the other resonances. 

Thus, the isotropic model is not accurate enough: in the following section we apply a model developed by Ahmad \etal for transverse isotropic cylinder.\cite{Ahmad00}\\

The Q-factor ($Q=f_0/\Delta f_0$) can be estimated from the half-power width $\Delta f_{0}$ of the resonance peak at the ZGV point. In order not to underestimate the Q-factor, the signal acquisition time window $\Theta$ must be larger than the inverse of the bandwidth. We performed a measurement of the out-of-plane displacement at epicenter with a signal acquisition time window $\Theta=4$ ms. We obtained thinner resonances and a Q-factor at L(0,2)-ZGV frequency of $4\times10^3$ ($\Delta f = 1.1$ kHz and $f_0=4.38$ MHz). For flexural ZGV resonances, the Q-factor are slightly lower and vary from $2\times10^3$ to $2.5\times10^3$.

%          Theoretical model
% ------------------------------------

\section{Transverse isotropic model}\label{sec:TImodel}

In the linear theory of elasticity, anisotropic media are described by the stiffness tensor $c_{ijkl}$ (with $i,~j,~k,~l$ = 1 to 3). Using the Voigt's notation, they are represented by a $6\times6$ symmetric matrix $c_{\alpha\beta}$ ($\alpha$, $\beta$ = 1 to 6). Given the Cartesian coordinates ($x_1$, $x_2$, $x_3$) with the $x_3$-axis parallel to the cylinder $z$-axis, the sample is supposed to be isotropic in the ($x_1$, $x_2$) plane. Such transverse isotropic medium is described by five independent elastic constants : $c_{11}$, $c_{13}$, $c_{33}$, $c_{44}$, and $c_{66}$. Other elastic constants are related to these coefficients: $c_{22} = c_{11}, \, c_{23} = c_{13}, \, c_{55} = c_{44}, \, c_{12} = c_{11} – 2c_{66}$ or vanish.

\subsection{Dispersion equation}

As mentioned by Honarvar \etal,\cite{Honarvar07} the displacement vector $\bm{u}(r,\theta,z,t)$ can be derived from three scalar potential functions $\varphi$, $\chi$, and $\psi$.\cite{Honarvar96,Pan06} A simple representation initiated by Buchwald~\cite{Buchwald61} and used by Ahmad and Rahman~\cite{Ahmad00} is the following, 
\begin{equation}
\bm{u} = \bm{\nabla}\varphi + \bm{\nabla}\times(\chi \hat{e}_z) + \left(\frac{\partial\psi}{\partial z}-\frac{\partial\varphi}{\partial z}\right)\hat{e}_z.
\label{eq:motion}
\end{equation}
In cylindrical coordinates ($r$, $\theta$, $z$), the above representation leads to the following displacement components,
\begin{eqnarray*}
(u_r,\,u_\theta,\,u_z) = 
\left(\frac{\partial\varphi}{\partial r}+\frac{1}{r}\frac{\partial\chi}{\partial\theta},\,
      \frac{1}{r}\frac{\partial\varphi}{\partial\theta}-\frac{\partial\chi}{\partial r},\,
      \frac{\partial\psi}{\partial z}\right).
\label{eq:displacement}
\end{eqnarray*}
Harmonic solutions are given by
\begin{eqnarray*}
\varphi_n &=& \Phi J_n(\beta r)\cos (n\theta)\exp [i(kz - \omega t)],\\
\psi_n    &=& \Psi J_n(\beta r)\cos (n\theta)\exp [i(kz - \omega t)],\\
\chi_n    &=&    X J_n(\beta r)\sin (n\theta)\exp [i(kz - \omega t)],
\label{eq:potential}
\end{eqnarray*}

where $J_n$ is the Bessel function of first kind of order $n$. They correspond to the superposition of plane waves of wavevector $\bm{k}=(\beta \cos\theta,\beta \sin\theta, k)$ which satisfy the propagation equation in a meridian plane. The radial component $\beta$ of the wavevector $\bm{k}$ must satisfy the \textit{Christoffel} equation
\begin{equation}
\left(c_{11}c_{44}\beta^4-E\beta^2+F\right)\left(c_{66}\beta^2+ c_{44}k^2-\rho\omega^2\right)=0.
\label{eq:christoffel}
\end{equation}
The first term correspond to waves polarized in the meridian plane ($x_2$, $x_3$) and the second term to pure shear wave ($x_1$). The coefficients $E$ and $F$ are given in Appendix [Eq.~\eqref{eq:EF}]. The above equations are similar to those obtained by Mirsky.\cite{Mirsky65a} Omitting the propagation term $\exp{i(kz-\omega t)}$ for simplicity, the three independent solutions are found to be
\begin{eqnarray*}
(\varphi_{1n},\,\psi_{1n},\,\chi_{1n}) &=& 
\Big(J_n(\beta_1 r)\cos(n\theta),\,
      q_1J_n(\beta_1 r)\cos(n\theta),\,
      0\Big),\\
(\varphi_{2n},\,\psi_{2n},\,\chi_{2n}) &=& 
\Big(q_2J_n(\beta_2r)\cos(n\theta),\,
         J_n(\beta_2r)\cos(n\theta),\,
         0\Big),\\       
(\varphi_{3n},\psi_{3n},\chi_{3n}) &=& 
\Big(0,\,0,\,J_n(\beta_3 r)\sin(n\theta)\Big),       
\end{eqnarray*}
where $q_1$ and $q_2$ are the potential amplitude ratios provided in Appendix Eq.~\eqref{eq:q1q2}. Hence, the displacements are as follows
%
%\begin{widetext}
\begin{eqnarray}\label{eq:disp}
(u_{1r},\,u_{1\theta},\,u_{1z})_n &=& 
\left(\beta_1J_n'(\beta_1 r)\cos(n\theta),\,
      -\left({n \over r }\right) J_n(\beta_1 r)\sin(n\theta),\,
       ikq_1J_n(\beta_1 r)\cos(n\theta)\right),\nonumber\\
(u_{2r},\,u_{2\theta},\,u_{2z})_n &=& 
 \left(q_2\beta_2J_n'(\beta_2 r)\cos(n\theta),\,
       -\left({n\over r}\right)q_2J_n(\beta_2 r)\sin(n\theta),\,
       ikJ_n(\beta_2 r)\cos(n\theta)\right),\\
(u_{3r},\,u_{3\theta},\,u_{3z})_n &=& 
\left(\left({n \over r}\right)J_n(\beta_3 r)\cos(n\theta),\,
     -\beta_3 J_n'(\beta_3 r)sin(n\theta),\,
       0\right).\nonumber
\end{eqnarray}
%\end{widetext}
%
The first two solutions correspond to a coupling between quasi-longitudinal and quasi-shear waves polarized in the meridian plane. The last solution having zero displacement component along the $z$-axis corresponds to a pure shear wave. These solutions must be combined with weighting factors $B_n, C_n, D_n$ to satisfy the boundary conditions at the free surface $r=a$. These conditions imply that normal stresses vanish: $\sigma_{rr}(a)=\sigma_{r\theta}(a)=\sigma_{rz}(a)=0$ and lead to the following homogeneous linear system 
\begin{equation}
\begin{bmatrix}
a_{11} & a_{12} & a_{13}\\ 
a_{21} & a_{22} & a_{23}\\ 
a_{31} & a_{32} & a_{33} 
\end{bmatrix}
\begin{bmatrix}
B_n\\ 
C_n\\ 
D_n\\  
\end{bmatrix}=
\begin{bmatrix}
0\\ 
0\\ 
0\\  
\end{bmatrix},
\label{eq:aijmat}
\end{equation}
where the matrix elements $a_{ij}$ are listed in the Appendix Eq.~\eqref{eq:aij_elements}. The dispersion equation results from the secular equation 
\begin{equation}
det(a_{ij})=0. 
\label{eq:secular}
\end{equation}
For $n=0$, Eq.~\eqref{eq:secular} splits into two parts: $a_{23}=0$, \ie, $(\beta_3 a)J_0^{''}(\beta_3 a)=J_0^{'}(\beta_3 a)$ similar to the equation given by Mirsky~\cite{Mirsky65a} for torsional modes T$(0,m)$ and $a_{11}a_{32}=a_{12}a_{31}$ corresponding to longitudinal modes L$(0,m)$.

\subsection{Cutoff frequencies}

As the wavenumber $k$ is equal to zero, the motion at the cutoff frequencies is independent of the axial coordinate $x_3$. This implies $k^2q_1=0$, $q_2=0$ and $a_{12}=a_{22}=0$. The dispersion equation [Eq.~\eqref{eq:secular}] for longitudinal and flexural modes in a solid cylinder at $k=0$ reduces to
\begin{equation}
\beta_2aJ_n^{'}(\beta_2a)(a_{13}a_{21}-a_{11}a_{23}) = 0,
\label{eq:DispCutoff}
\end{equation}
where the four $a_{ij}$ simplify as
\begin{eqnarray*}
a_{11} &=& c_{11}(\beta_1a)^2J''_n(\beta_1a)+c_{12} \big\{(\beta_1a)J'_n(\beta_1a)-
           n^2J_n(\beta_1a)\big\},\\
a_{13} &=& 2nc_{66}\big\{(\beta_3 a)J'_n(\beta_3 a)-J_n(\beta_3 a)\big\},\\
a_{21} &=& -2n\big\{(\beta_1 a)J'_n(\beta_1a)-J_n(\beta_1a)\big\},\\
a_{23} &=& -(\beta_3 a)^2J''_n(\beta_3 a) +(\beta_3 a)J'_n(\beta_3 a)-n^2 J_n(\beta_3 a).
\end{eqnarray*}

To find the cutoff frequencies, the above equation can be solved for $\omega_c(=2\pi f_c)$, for various values of $n$ and $m$. The first two roots $\beta_{1,2}$ are simplified using the reduced expressions of $\Delta=(c_{11}-c_{44})\rho\omega_c^2$, $E=(c_{11}+c_{44})\rho\omega_c^2$ and $F=\rho^2\omega_c^4$ [Eq.~\eqref{eq:beta123}] as $\beta_1=\omega_c\sqrt{\rho/c_{11}}$ and $\beta_2=\omega_c\sqrt{\rho/c_{44}}$. The third root reduces to $\beta_3=\omega_c\sqrt{\rho/c_{66}}$. The cutoff frequency equation [Eq.~\eqref{eq:DispCutoff}] is simplified as $a_{11}J_n^{'}(\beta_2a) = 0$ for L$(0,m)$.

\subsection{Dependence of ZGV frequencies on elastic constants}

\begin{figure*}[!ht]
\centering
\includegraphics[width=\textwidth]{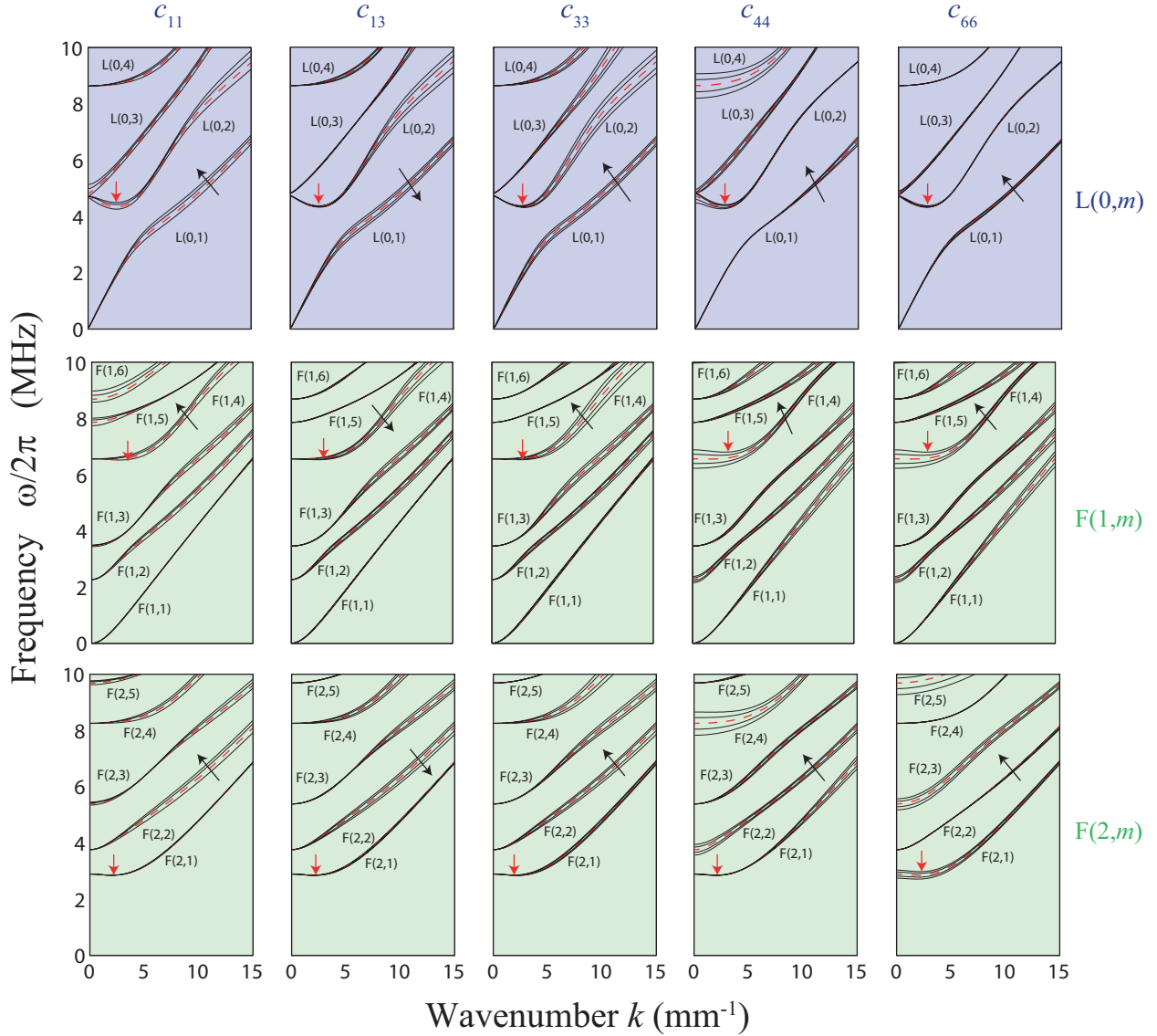}
\caption{Influence of the elastic constants on the dispersion curves of a transversely isotropic cylinder. From top to bottom: for longitudinal modes and the first two flexural mode families. Red arrows depict the ZGV frequencies and black arrows show increasing $c_{ij}$. (color online)}
\label{influence_cij}
\end{figure*}

A sensitivity analysis was performed to characterize the elastic constant influence on the dispersion curves and especially on ZGV frequencies. Figure \ref{influence_cij} displayed the longitudinal and the first two flexural mode families as a function of elastic constants. Each constant was successively varied by $\pm5\%$ and $\pm10\%$ around the following average values (in GPa): $c_{11}=240$, $c_{13}=110$, $c_{33}=250$, $c_{44}=70$, and $c_{66}=70$. The curves obtained for the average values are displayed in red dashed line. It appears that each ZGV resonance depends on specific constants. For example, the elastic constants which mainly affect the L(0,2)-ZGV frequency are $c_{11}$ and $c_{44}$. These elastic constants are proportional to the square of longitudinal and shear velocities in directions perpendicular to the cylinder axis $x_3$. While F(1,4)-ZGV frequency slightly depends on $c_{11}$ and more strongly on $c_{44}$ and $c_{66}$ which is proportional to the shear velocity in the direction perpendicular to the cylinder axis. When the values of $c_{11}$, $c_{13}$ and $c_{33}$ increase or when $c_{44}$, $c_{66}$ decrease, the F(1,4) minimum ZGV frequency disappears, \ie, the slope of the spectral line at the cutoff frequency becomes positive. Then, F(2,1)-ZGV frequency is almost independent of the first four elastic constants while it significantly depends on $c_{66}$.

% Discussion
% ------------------

\section{Results and discussion}\label{sec:results}

In this section, we identify all the resonance frequencies and accurately determine the five elastic constants. The theoretical dispersion curves obtained with the TI model~\cite{Ahmad00} were fitted with those obtained experimentally. Then, we compared isotropic and TI models to highlight the anisotropy of the stainless steel cylinder.

\subsection{Dispersion curves measurements}

\begin{figure*}[!ht]
\centering
\subfigure{\includegraphics[width=0.9\textwidth]{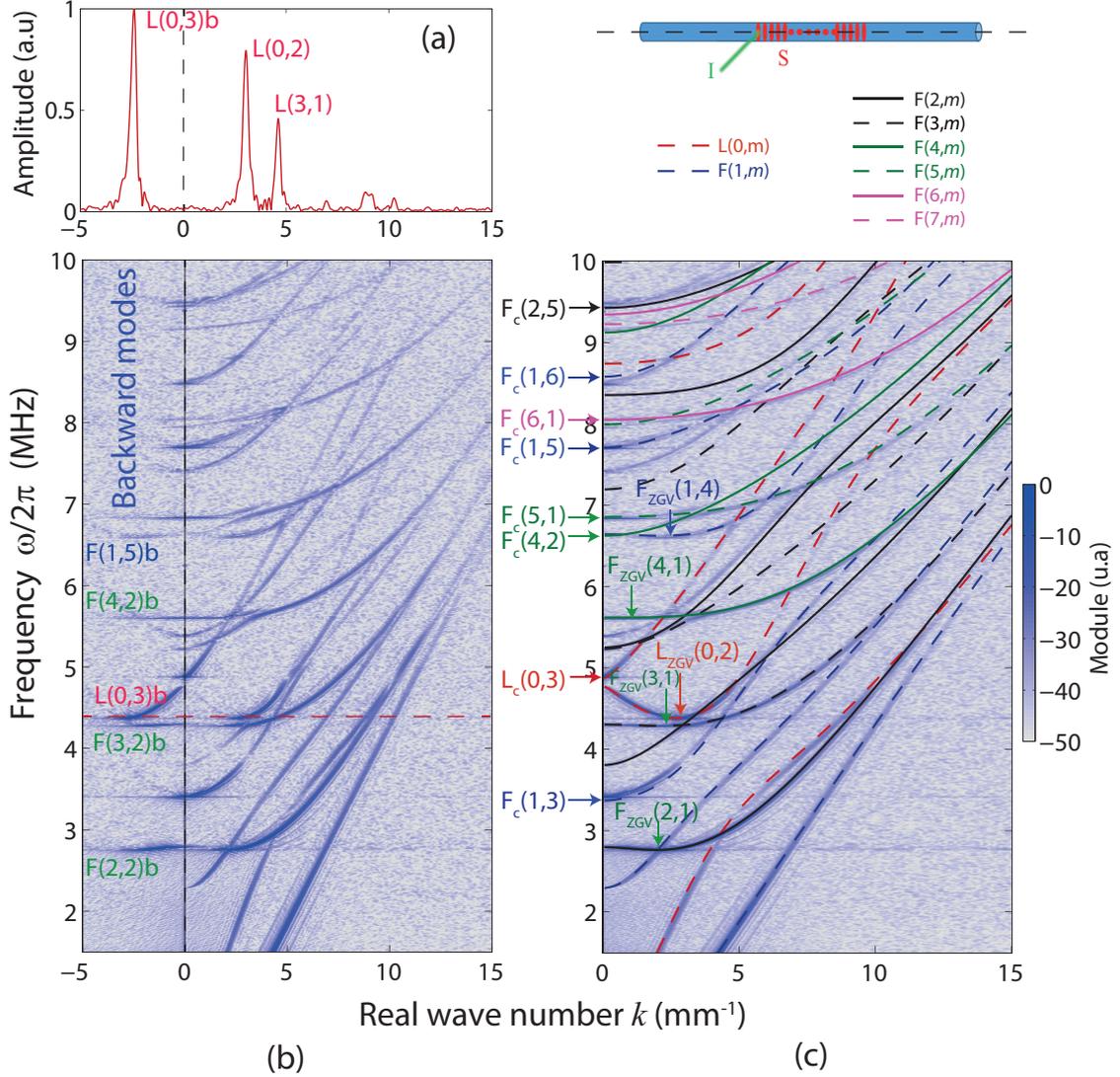}\label{exp_disp_curve_a}}
\subfigure{\label{exp_disp_curve_b}}\subfigure{\label{exp_disp_curve_c}}
\caption{a) Spatial Fourier transform of the normal displacement at 4.40 MHz illustrating backward propagation with a negative wavenumber. Each backward mode is coupled to the upper mode of the same family ($n$,$m+1$). To avoid confusion they are indicated by the index b. b) Experimental dispersion curves obtained for negative and positive wavenumbers. c) Negative dispersion curves $\omega(-k)$ were folded and added to the positive ones. The experimental dispersion curves were fitted with the TI model (the control parameters used are listed in the Tab.~\ref{tab:cij_TI}). (color online)}
\label{exp_disp_curve}
\end{figure*}

Experimental dispersion curves were measured with the laser ultrasound setup displayed in Fig.~\ref{setup}. The out-of-plane displacement was recorded with the interferometer, at a single point in the middle of the cylinder, while the line source was moved along the cylinder axis on 40 mm by 0.1 mm steps. For each source position, the normal displacement was recorded during 180 µs at a 50 MHz sampling frequency. 1024 signals were averaged to increase the signal to noise ratio. First, apodization (Hanning) windows were applied in both dimensions (time and distance) to avoid secondary lobes. Then, a 2D-Fourier transform was applied to the obtained B-scans. In order to observe the backward modes, the spatial Fourier transform was calculated for negative and positive wavenumbers ($k$) as shown in Fig.~\ref{exp_disp_curve_b}.\\ 

As previously discussed for Lamb modes in plates,\cite{Prada08b} cylindrical waveguides support backward-propagation with opposite group and phase velocities.\cite{Cui14} The counter propagative modes are also observed in the cylinder both for longitudinal and flexural modes and occur in the vicinity of $k=0$. For example, at the L(0,2)-ZGV frequency (4.34 MHz) the L(0,2) and L(0,3) modes having opposite wave vectors interfere. The power spectrum was computed at a frequency of 4.40 MHz, roughly higher than the L(0,2)-ZGV resonance frequency, for which the modes are propagative. Figure \ref{exp_disp_curve_a}, shows that this spectrum is composed of two main peaks. The peak at a negative value of $k$, similar to the larger one in the positive $k$ domain demonstrates the backward propagation in cylinders. Afterwards, the dispersion curves (backward region) for negative wavenumber $\omega(-k)$ was folded and added to positive ones [Fig.~\ref{exp_disp_curve_c}]. We can clearly identified five ZGV frequencies. One at 4.34 MHz corresponding to the L(0,2)-ZGV resonance and four others at 2.77, 4.28, 5.60 and 6.60 MHz corresponding to F(2,1), F(3,1), F(4,1) and F(1,4) flexural-ZGV resonances, respectively. The wavelength at these ZGV frequencies extend from $2.3$ mm to $3.2$ mm ($\lambda/d\sim 3-4$) except for F(1,4)-ZGV frequency which has a 14.3-mm wavelength ($\lambda/d\sim18.6$).\\

\begin{table*}[!ht]
\centering
\caption{\label{tab:cij_TI} Elastic constants $c_{ij}$ (or sound velocities) used in the TI model. $a$ and $\rho$ were measured.}
\begin{tabularx}{\textwidth}{YYYYYY}
\hline\hline
\multicolumn{6}{c}{\textbf{Stiffness constants (\giga\pascal)}}\\
\hline
$c_{11}$  & $c_{12}$ & $c_{13}$ & $c_{33}$ & $c_{44}$ & $c_{66}$\\
$239$     & $108$    & $110$    & $252$    & $72$     & $65$  \\
\end{tabularx}\\
\begin{tabularx}{\textwidth}{YYYYYY}
\hline
\multicolumn{5}{c}{\textbf{Sound velocities (\meter\per\second)}} & 
\textbf{Diameter (\micro\meter)}\\
\hline
$V_{L1}$  & $V_{S1}$ & $V_{L3}$ & $V_{S3}$ & $V_b$  & 2a \\
$5500$    & $2875$  & $5650$  & $3010$  & $4808$ & $775\pm2$\\
\end{tabularx}\\
\begin{tabularx}{\textwidth}{YYYYYY}
\hline
\multicolumn{2}{c}{\textbf{Young modulus (\giga\pascal)}} & 
\multicolumn{3}{c}{\textbf{Poisson's ratio}}  &
\textbf{Density (\kilogram\per\meter\cubed)} \\
\hline
$E_\perp$ & $E_{//}$ & $\nu_{12}$ & $\nu_{13}$ & $\nu_{31}$ & $\rho$    \\
$183$     & $172$    & $0.317$    & $0.316$    & $0.298$    & $7910\pm5$\\
\hline\hline
\end{tabularx}
\end{table*}

The theoretical dispersion relation $\omega(k)$ [Eq.~\eqref{eq:secular}] were calculated by solving the secular equation with the zero-finding algorithm.\cite{Seco12} Experimental dispersion curves were fitted with the TI model. Bulk modes proportional to stiffness constants $c_{33}$ and $c_{44}$ were eliminated by dividing the determinant by $(\rho\omega^2-c_{33}k^2)(\rho\omega^2-c_{44}k^2)$. We calculated longitudinal modes and the first seven families of flexural modes. The theoretical dispersion curves were fitted heuristically with five control parameters and with the diameter and the mass density previously measured. We can observed [Fig.~\ref{exp_disp_curve_c}] a good agreement between experimental dispersion curves and those predicted by the TI model. The parameters used for the TI model are listed in Tab.~\ref{tab:cij_TI}. The red and blue dashed lines correspond to the longitudinal and first flexural modes respectively, the green dashed lines indicated the flexural modes with $n\geqslant2$. Although most modes are well fitted by the theory, a few experimental branches are not explained. This may be due to a small remaining error on the estimated constants or to a slight discrepancy with transverse anisotropy.

\subsection{Elastic parameters}

We now explain the physical meaning of the five sound velocities implicitly involved in the TI model. Stiffness constants $c_{11}$ and $c_{33}$ determine the longitudinal (L) ultrasound velocities in perpendicular ($V_{L1}$) and parallel ($V_{L3}$) directions with respect to the cylinder axis $x_3$
\begin{equation}
c_{11} = \rho V_{L1}^2, \, c_{33} = \rho V_{L3}^2.
\label{eq:c11c33}
\end{equation}
Stiffness constants $c_{44}$ and $c_{66}=(c_{11}-c_{12})/2$ determine the shear (S) velocities in directions perpendicular to the cylinder axis with polarizations either parallel ($V_{S1}$) or perpendicular ($V_{S3}$) to $x_3$-axis
\begin{equation}
c_{44} = \rho V_{S1}^2, \, c_{66} = \rho V_{S3}^2.
\label{eq:c44c66}
\end{equation}
The remaining stiffness constant $c_{13}$ should be adjusted manually. Hence, two Young's modulus can be defined: $E_{//}=E_{33}$ for a stress parallel to the cylinder axis and $E_{\perp}=E_{11}=E_{22}$ for a stress perpendicular to the cylinder axis. They are given by 
\begin{equation}
E_{//}  = \frac{c^2}{c_{11}-c_{66}}, \, E_\perp = \frac{4c_{66}c^2}{c^2+c_{33}c_{66}},
\label{eq:ETP}
\end{equation}
where $c^2=c_{33}(c_{11}-c_{66})-c_{13}^2$ is the determinant of the sub-matrix of elastic constants $c_{ij}$. This effective stiffness constant $c$ is proportional to the bar velocity of the waves propagating along the cylinder axis: $V_b = (E_{//}/\rho)^{1/2}$ (see Mirsky~\cite{Mirsky65a} Eq.~44). Three Poisson's ratios~\cite{Tokmakova08,Ballato10} can be defined, $\nu_{12}$ and $\nu_{31}$ for longitudinal extension in the basal plane and $\nu_{13}$ along the 6-fold axis
\begin{equation*}
\nu_{12} = \frac{c_{12}c_{33}-c_{13}^2}{c^2+c_{33}c_{66}},\,
\nu_{31} = \frac{2c_{66}c_{13}}{c^2+c_{33}c_{66}},
\nu_{13} = \frac{c_{13}}{c_{11}+c_{12}}.\, 
\label{eq:nuTP}
\end{equation*}

\subsection{Comparison between isotropic and TI model}

\begin{figure*}[!ht]
\subfigure{\includegraphics[width=\textwidth]{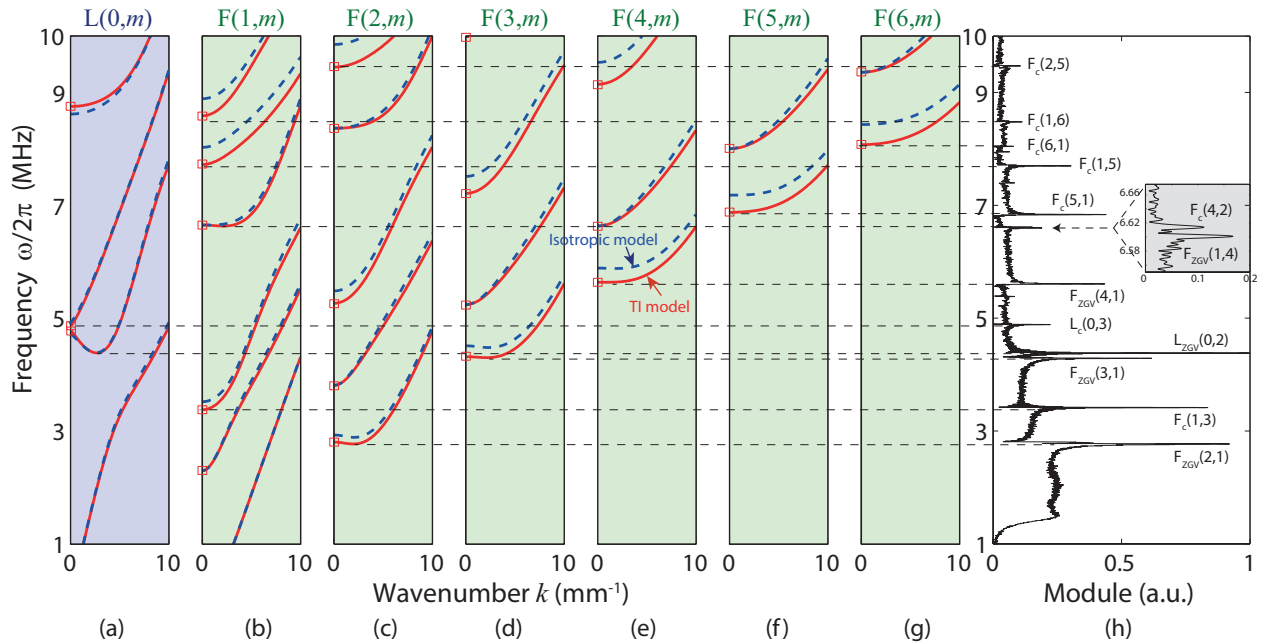}\label{TI_ISO_cmp1}}
\subfigure{\label{TI_ISO_cmp2}}\subfigure{\label{TI_ISO_cmp3}}\subfigure{\label{TI_ISO_cmp4}}
\subfigure{\label{TI_ISO_cmp5}}\subfigure{\label{TI_ISO_cmp6}}\subfigure{\label{TI_ISO_cmp7}}\subfigure{\label{TI_ISO_cmp8}}
\caption{Comparison of isotropic model (blue dashed lines) with TI model (red lines). a) Longitudinal mode L$(0,m)$. b-g) Flexural modes F$(n,m)$ for $n$ extend from $1$ to $6$. h) Frequency spectrum recorded at epicenter. Inset: two resonance frequencies are observed corresponding to F(1,4)-ZGV and F$_c$(4,2), respectively. The red square ($\textcolor{red}{\square}$) indicate the cutoff frequencies calculated with [Eq.~\eqref{eq:DispCutoff}]. (color online)}
\label{TI_ISO_cmp}
\end{figure*}

Theoretical dispersion curves obtained with isotropic (blue dashed lines) and transversely isotropic (red lines) models are plotted in Fig.~\ref{TI_ISO_cmp}. The longitudinal modes are displayed in [Fig.~\ref{TI_ISO_cmp1}] and the first six flexural mode families [Figures~\ref{TI_ISO_cmp2}-\ref{TI_ISO_cmp7}].\\

The frequency spectrum recorded at the epicenter [Fig.~\ref{TI_ISO_cmp8}]. This representation allows to distinguish the different propagation modes and to precisely identify each resonance peak. It obviously appears that the isotropic model is not appropriate to interpret experimental results and clearly confirms the transverse isotropic nature of the cylinder. Five ZGV resonances, F(2,1), F(3,1), L(0,2), F(4,1), F(1,4), and eight thickness resonances, F(1,3), L(0,3), F(4,2), F(5,1), F(1,5), F(6,1), F(1,6) and F(2,5), can be identified from the dispersion curves calculated with the TI model. Both ZGV and cutoff frequencies are estimated with a relative error less than $0.6\%$, except for F$_c$(1,3) which was estimated with an error close to $1.1\%$.

\subsection{Displacements at cutoff frequencies}

\begin{figure}[!ht]
\includegraphics[width=0.75\columnwidth]{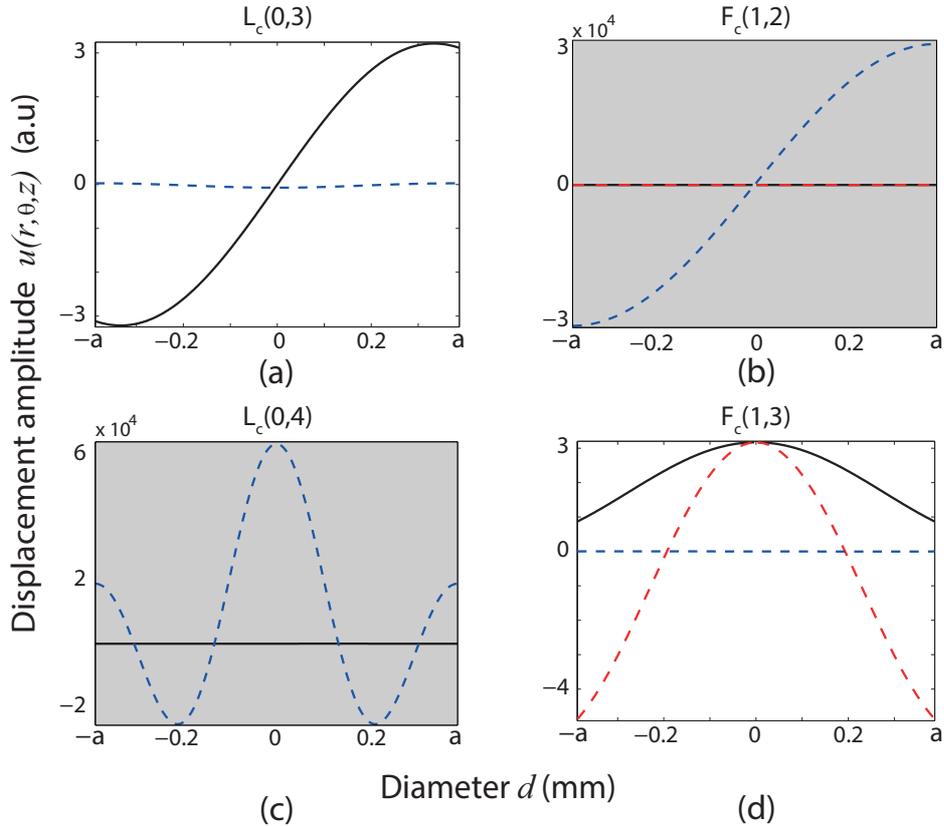}
\caption{Azimuthal displacement produced at a) L(0,3), b) F(1,2), c) L(0,4) and d) F(1,3) cutoff frequencies. Black (---), red dash (\textcolor{red}{- - -}) and blue (\textcolor{blue}{- - -}) dash lines indicate $u_r$, $u_\theta$ and $u_z$, respectively. (color online)}
\label{cutoff_disp}
\end{figure}

In order to explain why several thickness resonances (at $k=0$) are observed while others are not, we analysed the displacements generated inside the cylinder. To this end, we calculated $u_r$, $u_\theta$ and $u_z$, using Eq.~\eqref{eq:disp}. The radial and axial displacements ($u_r$) and ($u_z$) were calculated at $\theta=0$  and the azimuthal displacement ($u_\theta$) at $\theta_{max}=\pi/2n$. They are displayed in Fig.~\ref{cutoff_disp} at the cutoff frequencies of the modes L(0,3), L(0,4), F(1,2) and F(1,3) which respectively correspond to the frequencies 4.834, 8.673, 2.276 and 3.343 MHz. For L$_c$(0,3) and F$_c$(1,3)  thickness resonances,  the radial displacement at the  cylinder surface $r=a$ is significant. Conversely, for the modes L$_c$(0,4) and F$_c$(1,2) the radial displacement vanishes for $r=a$, but have essentially a large displacement along z-axis. This explains why, at these cutoff frequencies, no resonances are observed.

% Conclusion
% -----------------

\section{Conclusion}

In summary, elastic guided waves propagating in a stainless steel cylinder of millimetric diameter were investigated by laser ultrasonic techniques. Using a laser line source perpendicular to the cylinder axis, ZGV resonances were observed both for longitudinal and flexural modes. One of the major interests is to identify each resonance to characterize elastic properties of materials. We demonstrated that isotropic model was not relevant to describe the behaviour of elastic guided waves in a steel rod. A transverse isotropic model allowed us to describe the propagation and to calculate dispersion curves $\omega(k)$. The five elastic constants were estimated from experimental dispersion curves measured by the LU technique. A good agreement between theoretical and experimental dispersion curves was obtained both for longitudinal and flexural modes.

In the future, it would be useful to build a numerical inversion procedure to improve the accuracy of the material parameter estimation. Furthermore, it will be interesting to investigate the relationship between ZGV modes and elastic constants. This could be done by calculating the second order derivatives of the dispersion curves at cutoff frequencies to determine the existence of backward modes. Finally, it will be interesting to conduct other studies in materials with different types of anisotropy. This technique can be also applied in the sub-gigahertz range with a laser line source of width comparable to the cylinder diameter to study micrometric fibers.

%        Acknowledgements
% ------------------------------------
\begin{acknowledgements}
JL, DR and CP was supported by LABEX WIFI (Laboratory of Excellence ANR-10-LABX-24) within the French Program ``Investments for the Future'' under reference ANR-10-IDEX-0001-02 PSL$^{\ast}$.
\end{acknowledgements}

%       Appendix
% -----------------------
\appendix*
\section{}\label{sec:A1}

The scalar functions $\varphi$, $\chi$ and $\psi$ satisfy the following wave motion equations
\begin{eqnarray*}\label{eq:motion3}
c_{11}\left(\bm{\nabla^2}-\frac{\partial^2}{\partial z^2} \right )\varphi +c_{44}\frac{\partial^2\varphi}{\partial z^2}
+(c_{13}+c_{44})\frac{\partial^2\psi}{\partial z^2} &=& \rho\frac{\partial^2\varphi}{\partial t^2},\\
(c_{13}+c_{44})\left(\bm{\nabla^2}-\frac{\partial^2}{\partial z^2} \right )\varphi +c_{44}\left(\bm{\nabla^2}-\frac{\partial^2}{\partial z^2} \right )\psi\\
+c_{33}\frac{\partial^2\psi}{\partial z^2}&=&\rho\frac{\partial^2\psi}{\partial t^2},\\
c_{66}\left(\bm{\nabla^2}-\frac{\partial^2}{\partial z^2} \right )\chi + c_{44}\frac{\partial^2\chi}{\partial z^2}&=&\rho\frac{\partial^2\chi}{\partial t^2}.
\end{eqnarray*}
The longitudinal L (represented by $\varphi$) and the vertically polarized quasi-transverse SV (represented by $\psi$) waves are coupled, whereas the pure transverse wave SH (represented by $\chi$) is decoupled from the others. The dispersion equation of torsional, longitudinal and flexural guided modes in a transversely isotropic cylinder with free boundary conditions results in the vanishing of the determinant of a $3\times3$ matrix $a_{ij}$
\begin{equation*}
%\label{eq:detTLF}
det(a_{ij})=
\left\{\begin{array}{ll}
a_{23} = 0 &\text{for T$(0,m)$},\\
a_{11}a_{32} - a_{12}a_{31} = 0 &\text{for L$(0,m)$},\\
a_{11}(a_{22}a_{33} - a_{23}a_{32})-\\
a_{12}(a_{21}a_{33} - a_{23}a_{31})+\\ 
a_{13}(a_{21}a_{32} - a_{22}a_{31}) = 0  &\text{for F$(n>1,m)$},
\end{array}\right.
\end{equation*}
where the matrix elements are
\begin{eqnarray}\label{eq:aij_elements}
a_{11} &=& c_{11}(\beta_1a)^2J_n^{''}(\beta_1a)+\nonumber\\
        && c_{12}\left \{(\beta_1a)J_n^{'}(\beta_1a)-n^2J_n(\beta_1a) \right \} -\nonumber\\
        && c_{13}q_1(ka)^2J_n(\beta_1a),\nonumber\\
a_{12} &=& c_{11}q_2(\beta_2a)^2J_n^{''}(\beta_2a)+\nonumber\\
        && c_{12}q_2\left \{(\beta_2a)J_n^{'}(\beta_2a)-n^2J_n(\beta_2a) \right \} -\nonumber\\
        && c_{13}(ka)^2J_n(\beta_2a),\nonumber\\
a_{13} &=& 2nc_{66}\left \{(\beta_3 a)J_n^{'}(\beta_3 a)-J_n(\beta_3 a)\right \},\nonumber\\       
a_{21} &=& -2n\left \{(\beta_1a)J_n^{'}(\beta_1a)-J_n(\beta_1a)\right \},\\
a_{22} &=& -2nq_2\left \{(\beta_2a)J_n^{'}(\beta_2a)-J_n(\beta_2a)\right \},\nonumber\\
a_{23} &=& -(\beta_3 a)^2J_n^{''}(\beta_3 a)+(\beta_3 a)J_n^{'}(\beta_3 a)-n^2J_n(\beta_3 a),\nonumber\\           
a_{31} &=& (1+q_1)(\beta_1a)J_n^{'}(\beta_1a),\nonumber\\
a_{32} &=& (1+q_2)(\beta_2a)J_n^{'}(\beta_2a),\nonumber\\    
a_{33} &=& nJ_n(\beta_3 a),\nonumber
\end{eqnarray}
respectively. The first and second derivatives of the Bessel function of the first kind of order $n$ can be expressed in terms of $J_{n-1}$, $J_n$, $J_{n+1}$ etc. by using recurrence relations.

The coefficients $E$ and $F$ used in the \textit{Christoffel} Equation [Eq.~\ref{eq:christoffel}] are defined as
\begin{eqnarray}\label{eq:EF}
E &=& (c_{13}+c_{44})^2k^2+c_{44}(\rho\omega^2-c_{44}k^2)+c_{11}(\rho\omega^2-c_{33}k^2)\nonumber\\
  &=& E_1^2+c_{44}F_1+c_{11}F_2,\\
F &=& (\rho\omega^2-c_{44}k^2)(\rho\omega^2-c_{33}k^2)=F_1F_2.\nonumber
\end{eqnarray}
Finally, the three roots ($\beta$) of \textit{Christoffel} equation are
\begin{equation}\label{eq:beta123}
\beta_{1,2} = \sqrt{\frac{E\mp\Delta}{2c_{11}c_{44}}} \quad\text{and}\quad
\beta_3     = \sqrt{\frac{F_1}{c_{66}}},
\end{equation}
where $\Delta = \sqrt{E^2-4c_{11}c_{44}F}$. At last, the amplitude ratio $q_1$ and $q_2$ are given by
\begin{equation}
q_1 = \frac{F_1-c_{11}\beta_1^2}{kE_1},\quad
q_2 = \frac{kE_1}{F_1-c_{11}\beta_2^2}.
\label{eq:q1q2}
\end{equation}

% Bibliographie
% -------------------
%\bibliographystyle{custom}
%\bibliography{Biblio/BiblioZGV}

\begin{thebibliography}{10}

\bibitem{Gazis59}
D.~C. Gazis, Three-dimensional investigation of the propagation of waves in
  hollow circular cylinders. {I}. {A}nalytical foundation. {II}. {N}umerical
  results, \emph{\href{http://dx.doi.org/10.1121/1.1907753}{J. Acoust. Soc.
  Am.}} \textbf{31}, 568--578 (1959).

\bibitem{Mindlin60}
R.~Mindlin and H.~McNiven, Axially symmetric waves in elastic rods,
  \emph{\href{http://dx.doi.org/10.1115/1.3643889}{J. Appl. Mech.}}
  \textbf{27}, 145--151 (1960).

\bibitem{Meeker64}
T.~Meeker and A.~Meitzler, Guided wave propagation in elongated cylinders and
  plates,
  \emph{\href{http://dx.doi.org/10.1016/B978-1-4832-2857-0.50008-3}{Phys.
  Acoust.}} \textbf{1}, 111--167 (1964).

\bibitem{Hudson43}
G.~E. Hudson, Dispersion of elastic waves in solid circular cylinders,
  \emph{\href{http://dx.doi.org/10.1103/PhysRev.63.46}{Phys. Rev.}}
  \textbf{63}, 46 (1943).

\bibitem{Davies48}
R.~Davies, A critical study of the hopkinson pressure bar,
  \emph{\href{http://dx.doi.org/10.1098/rsta.1948.0001}{Phil. Trans. Roy. Soc.
  London. A.}} pages 375--457 (1948).

\bibitem{Pao60}
Y.-H. Pao and R.~Mindlin, Dispersion of flexural waves in an elastic, circular
  cylinder, \emph{\href{http://dx.doi.org/10.1115/1.3644033}{J. Appl. Mech.}}
  \textbf{27}, 513--520 (1960).

\bibitem{Pao62}
Y.-H. Pao, The dispersion of flexural waves in an elastic circular cylinder,
  part 2, \emph{\href{http://dx.doi.org/10.1115/1.3636498}{J. Appl. Mech.}}
  \textbf{29}, 61--64 (1962).

\bibitem{Onoe62}
M.~Onoe, H.~McNiven, and R.~Mindlin, Dispersion of axially symmetric waves in
  elastic rods, \emph{\href{http://dx.doi.org/10.1115/1.3640661}{J. Appl.
  Mech.}} \textbf{29}, 729--734 (1962).

\bibitem{Zemanek72}
J.~Zemanek~Jr, An experimental and theoretical investigation of elastic wave
  propagation in a cylinder,
  \emph{\href{http://dx.doi.org/10.1121/1.1912838}{J. Acoust. Soc. Am.}}
  \textbf{51}, 265--283 (1972).

\bibitem{Mason99}
T.~Mason, Variation in the dispersion of axisymmetric waves in infinite
  circular rods with crystallographic wire texture,
  \emph{\href{http://dx.doi.org/10.1121/1.427160}{J. Acoust. Soc. Am.}}
  \textbf{106}, 1262--1270 (1999).

\bibitem{Payton83}
R.~G. Payton, \emph{Elastic wave propagation in transversely isotropic media},
  volume~4, Springer Science \& Business Media (1983).

\bibitem{Morse54}
R.~Morse, Compressional waves along an anisotropic circular cylinder having
  hexagonal symmetry, \emph{\href{http://dx.doi.org/10.1121/1.1907440}{J.
  Acoust. Soc. Am.}} \textbf{26}, 1018--1021 (1954).

\bibitem{Mirsky64}
I.~Mirsky, Axisymmetry vibrations of orthotropic cylinders,
  \emph{\href{http://dx.doi.org/10.1121/1.1919329}{J. Acoust. Soc. Am.}}
  \textbf{36}, 2106--2112 (1964).

\bibitem{Mirsky65a}
I.~Mirsky, Wave propagation in transversely isotropic circular cylinders, part
  {I}: {T}heory, part {II}: {N}umerical results,
  \emph{\href{http://dx.doi.org/10.1121/1.1909508}{J. Acoust. Soc. Am.}}
  \textbf{37}, 1016--1026 (1965).

\bibitem{Eliot68}
F.~C. Eliot and G.~Mott, Elastic waves propagation in circular cylinders having
  hexagonal crystal symmetry,
  \emph{\href{http://dx.doi.org/10.1121/1.1911097}{J. Acoust. Soc. Am.}}
  \textbf{44}, 423--430 (1968).

\bibitem{Mcniven71}
H.~McNiven and Y.~Mengi, Dispersion of waves in transversely isotropic rods,
  \emph{\href{http://dx.doi.org/10.1121/1.1912321}{J. Acoust. Soc. Am.}}
  \textbf{49}, 229--236 (1971).

\bibitem{Nagy95}
P.~B. Nagy, Longitudinal guided wave propagation in a transversely isotropic
  rod immersed in fluid, \emph{\href{http://dx.doi.org/10.1121/1.413702}{J.
  Acoust. Soc. Am.}} \textbf{98}, 454--457 (1995).

\bibitem{Berliner96a}
M.~J. Berliner and R.~Solecki, Wave propagation in fluid-loaded, transversely
  isotropic cylinders, part {I}. {A}nalytical formulation, part {II}.
  {N}umerical results, \emph{\href{http://dx.doi.org/10.1121/1.415365}{J.
  Acoust. Soc. Am.}} \textbf{99}, 1841--1853 (1996).

\bibitem{Niklasson98}
A.~J. Niklasson and S.~K. Datta, Scattering by an infinite transversely
  isotropic cylinder in a transversely isotropic medium,
  \emph{\href{http://dx.doi.org/10.1016/S0165-2125(97)00038-3}{Wave Motion}}
  \textbf{27}, 169--185 (1998).

\bibitem{Ahmad00}
F.~Ahmad and A.~Rahman, Acoustic scattering by transversely isotropic
  cylinders, \emph{\href{http://dx.doi.org/10.1016/S0020-7225(99)00031-2}{Int.
  J. Eng. Sci.}} \textbf{38}, 325--335 (2000).

\bibitem{Buchwald61}
V.~Buchwald, Rayleigh waves in transversely isotropic media,
  \emph{\href{http://dx.doi.org/10.1093/qjmam/14.3.293}{Q. J. Mech. Appl.
  Math.}} \textbf{14}, 293--317 (1961).

\bibitem{Ahmad01}
F.~Ahmad, Guided waves in a transversely isotropic cylinder immersed in a
  fluid, \emph{\href{http://dx.doi.org/10.1121/1.1348299}{J. Acoust. Soc. Am.}}
  \textbf{109}, 886--890 (2001).

\bibitem{Honarvar96}
F.~Honarvar and A.~Sinclair, Acoustic wave scattering from transversely
  isotropic cylinders, \emph{\href{http://dx.doi.org/10.1121/1.415868}{J.
  Acoust. Soc. Am.}} \textbf{100}, 57--63 (1996).

\bibitem{Honarvar07}
F.~Honarvar, E.~Enjilela, A.~N. Sinclair, and S.~A. Mirnezami, Wave propagation
  in transversely isotropic cylinders,
  \emph{\href{http://dx.doi.org/10.1016/j.ijsolstr.2006.12.029}{Int. J. Sol.
  Struct.}} \textbf{44}, 5236--5246 (2007).

\bibitem{Honarvar09}
F.~Honarvar, E.~Enjilela, and A.~N. Sinclair, Asymmetric and axisymmetric
  vibrations of finite transversely isotropic circular cylinders,
  \emph{\href{http://dx.doi.org/10.1134/S1063771009060037}{Acoust. Phys.}}
  \textbf{55}, 708--714 (2009).

\bibitem{Norris10}
A.~N. Norris and A.~Shuvalov, Wave impedance matrices for cylindrically
  anisotropic radially inhomogeneous elastic solids,
  \emph{\href{http://dx.doi.org/10.1093/qjmam/hbq010}{Q. J. Mech. Appli.
  Math.}} \textbf{63}, 401--435 (2010).

\bibitem{Norris13}
A.~N. Norris, A.~J. Nagy, and F.~A. Amirkulova, Stable methods to solve the
  impedance matrix for radially inhomogeneous cylindrically anisotropic
  structures, \emph{\href{http://dx.doi.org/10.1016/j.jsv.2012.12.016}{J. Sound
  Vib.}} \textbf{332}, 2520--2531 (2013).

\bibitem{Chitikireddy11}
R.~Chitikireddy, S.~K. Datta, A.~H. Shah, and H.~Bai, Transient thermoelastic
  waves in an anisotropic hollow cylinder due to localized heating,
  \emph{\href{http://dx.doi.org/10.1016/j.ijsolstr.2011.06.023}{Int. J. Solids
  Struct.}} \textbf{48}, 3063--3074 (2011).

\bibitem{Hutchins89}
D.~A. Hutchins, K.~Lundgren, and S.~B. Palmer, A laser study of transient
  {L}amb waves in thin materials,
  \emph{\href{http://dx.doi.org/10.1121/1.397981}{J. Acoust. Soc. Am.}}
  \textbf{85}, 1441 (1989).

\bibitem{Gsell04}
D.~Gsell and J.~Dual, Non-destructive evaluation of elastic material properties
  in anisotropic circular cylindrical structures,
  \emph{\href{http://dx.doi.org/10.1016/j.ultras.2004.02.026}{Ultrasonics}}
  \textbf{43}, 123--132 (2004).

\bibitem{Prada05a}
C.~Prada, O.~Balogun, and T.~Murray, Laser based ultrasonic generation and
  detection of {Z}ero-{G}roup {V}elocity {L}amb waves in thin plates,
  \emph{\href{http://dx.doi.org/10.1063/1.2128063}{Appl. Phys. Lett.}}
  \textbf{87}, 194109 (2005).

\bibitem{Prada08b}
C.~Prada, D.~Clorennec, and D.~Royer, Local vibration of an elastic plate and
  {Z}ero-{G}roup {V}elocity {L}amb modes,
  \emph{\href{http://dx.doi.org/10.1121/1.2918543}{J. Acoust. Soc. Am.}}
  \textbf{124}, 203 (2008).

\bibitem{Clorennec07}
D.~Clorennec, C.~Prada, and D.~Royer, Local and noncontact measurements of bulk
  acoustic wave velocities in thin isotropic plates and shells using zero group
  velocity {L}amb modes, \emph{\href{http://dx.doi.org/10.1063/1.2434824}{J.
  Appl. Phys.}} \textbf{101}, 034908 (2007).

\bibitem{Ces12}
M.~C{\`e}s, D.~Royer, and C.~Prada, Characterization of mechanical properties
  of a hollow cylinder with zero group velocity {L}amb modes,
  \emph{\href{http://dx.doi.org/10.1121/1.4726033}{J. Acoust. Soc. Am.}}
  \textbf{132}, 180 (2012).

\bibitem{Clorennec02}
D.~Clorennec, D.~Royer, and H.~Walaszek, Nondestructive evaluation of
  cylindrical parts using laser ultrasonics,
  \emph{\href{http://dx.doi.org/10.1016/S0041-624X(02)00210-X}{Ultrasonics}}
  \textbf{40}, 783--789 (2002).

\bibitem{Mounier14a}
D.~Mounier, C.~Poil{\^a}ne, H.~Khelfa, and P.~Picart, Sub-gigahertz laser
  resonant ultrasound spectroscopy for the evaluation of elastic properties of
  micrometric fibers,
  \emph{\href{http://dx.doi.org/10.1016/j.ultras.2013.06.014}{Ultrasonics}}
  \textbf{54}, 259--267 (2014).

\bibitem{Mounier14b}
H.~Khelfa, D.~Mounier, C.~Poil{\^a}ne, and P.~Picart, Evidence of guided
  acoustic waves propagating along a micrometric fiber,
  \emph{\href{http://dx.doi.org/10.1063/1.4899195}{Appl. Phys. Lett.}}
  \textbf{105}, 161906 (2014).

\bibitem{Seco12}
F.~Seco and A.~R. Jim{\'e}nez, \emph{Modelling the generation and propagation
  of ultrasonic signals in cylindrical waveguides}, Ultrasonic Waves, Dr Santos
  (Ed.), ISBN: 978-953-51-0201-4, InTech (2012).

\bibitem{Pan06}
Y.~Pan, M.~Perton, C.~Rossignol, and B.~Audoin, The transient response of a
  transversely isotropic cylinder under a laser point source impact,
  \emph{\href{http://dx.doi.org/10.1016/j.ultras.2006.05.180}{Ultrasonics}}
  \textbf{44}, e823--e827 (2006).

\bibitem{Cui14}
H.~Cui, W.~Lin, H.~Zhang, X.~Wang, and J.~Trevelyan, Characteristics of group
  velocities of backward waves in a hollow cylinder,
  \emph{\href{http://dx.doi.org/10.1121/1.4872297}{J. Acoust. Soc. Am.}}
  \textbf{135}, 3398--3408 (2014).

\bibitem{Tokmakova08}
S.~Tokmakova, Anisotropy of {P}oisson's ratio in transversely isotropic rocks,
  \href{http://dx.doi.org/10.1063/1.2956246}{in \emph{AIP Conf. Proc.}}, volume
  1022, pages 413--416 (2008).

\bibitem{Ballato10}
A.~Ballato, Poisson's ratios of auxetic and other technological materials,
  \emph{\href{http://dx.doi.org/10.1109/TUFFC.2010.1372}{IEEE Trans. Ultra.
  Ferro. Freq. Cont.}} \textbf{57}, 7--15 (2010).

\end{thebibliography}

% List of figures and tables
% ----------------------------
\newpage
\listoffigures
\listoftables

\end{document}